\documentclass[12pt]{article}

\usepackage{graphicx}
\usepackage{color}
\usepackage{mathptmx}      
\usepackage{amssymb}
\usepackage{amsmath}
\usepackage{ifsym}
\usepackage{sidecap}
\usepackage[colorlinks, linkcolor=blue, citecolor=blue, urlcolor=blue]{hyperref}

\usepackage{amssymb}

\begin{document}

\noindent
{\Large Coherence loss in stroboscopic radar ranging}

\noindent
{\Large in the problem of asteroid size estimation\footnote{\sl Preprint submitted to Acta Astronautica}}


\bigskip
\bigskip

\noindent
{\large V. D. Zakharchenko\footnote{\sl E-mail address: zakharchenko\_vd@mail.ru},  I. G. Kovalenko\footnote{\sl E-mail address: ilya.g.kovalenko@gmail.com},  V. Yu. Ryzhkov\footnote{\sl E-mail address:ozzy-74@mail.ru}
}

\bigskip
\noindent
{\it Volgograd State University,
         Universitetskij Pr., 100,  Volgograd 400062, Russia
}


\begin{abstract}
We consider the problem of coherence loss in a stroboscopic high resolution radar ranging due to phase instability of the probing and reference radio signals. Requirements to the coherence of reference generators in stroboscopic signal processing system are formulated. The results of statistical modeling are presented. \end{abstract}



\section{Preamble}\label{Preamble}

Development of the methods that allow improving accuracy of determining the asteroid sizes  (i.e. whether they measuredozens or hundreds meters in diameter) is important for correct estimateof damage they can cause (either regional or global catastrophes, respectively). At the same time this research can be interesting for specialists who study shapes and the surface geometry of small bodies of the Solar system.

In our previous works \cite{Zakharchenko15, ZakhKovpatent14} we proposed the method to estimate sizes of passive cosmic objects which method utilizes the radiolocation probing by ultra-high-resolving nanosecond signals for obtaining radar signatures. The method involves radio pulse strobing of reflected ultra-high-resolving signals from the surface of the cosmic object. The complete coherence of the probing and reflected signals is an essential condition of the method. However such a condition corresponds to idealized case when no phase instabilities exist in the signal processing system. The real sources of reference oscillations have nonzero instant instability of frequency which leads to loss of coherence at large signal lags (large distances). This factor restricts performance of coherent processing methods and leads to reduction of signal-to-noise ratio at the output of a stroboscopic system.

In the analysis of time scale transformation of broadband radio signals \cite{Zakharchenko15, ZakhKovpatent14} the complete coherence of carrier frequencies of the measured and the reference oscillations is commonly assumed. Such a concept corresponds to an absence of phase instabilities in the signal processing system. In real devices this condition can be broken due to deviations of  reference generator frequency and phase, instabilities of delays in a signal path and other factors.  These factors restrict performance of coherent processing methods and lead to reduction of signal-to-noise ratio at the output of a stroboscopic system.

Let us consider the influence of phase instability of the carrier frequencies of the measured and the reference signals on statistical characteristics of the transformed signal in stroboscopic processing. We will describe  the loss of coherence by a random process $\theta(t)$,  i.e. by a fluctuation component of a phase difference between the received and the strobe radio signals. The statistical characteristics of the phase difference  $\theta(t)$   are considered to be known. In the analysis we will assume that the coefficient of spectral transformation $N$  is large enough to use asymptotic estimates.

\begin{figure}
  \includegraphics[width=1\textwidth]{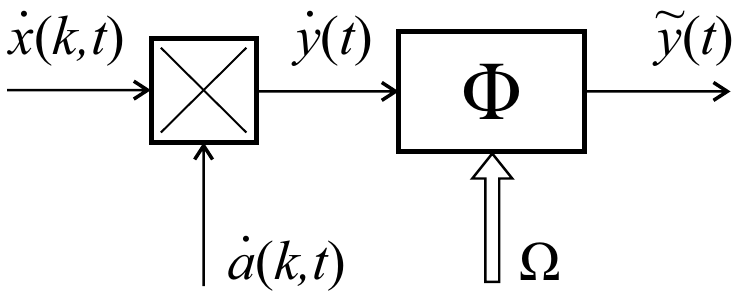}
\caption{The model of stroboscopic processing of a Doppler signal.}
\label{fig:1}
\end{figure}

The model of stroboscopic processing of the reflected signals (Fig.~\ref{fig:1}) differs from one considered in the work \cite{Zakharchenko15} by the low-pass filter being replaced with the tracking filter which adaptively tunes to differential frequency of carriers $\Omega=2\omega_0 V_r/c$   where $\omega_0$  is the carrier frequency of the probing signal, $V_r$  is the radial velocity of an asteroid.

For an exact determination of differential frequency $\Omega$, the radial velocity $V_r$  of an asteroid has to be measured independently using narrow-band methods based on center of mass of the Doppler signal spectrum. One of such  effective methods is the method of real-time assessment of radial velocity by means of fractional differentiation of a Doppler signal considered in the previous work of the authors \cite{ZakhKov14}.

\section{Modeling of coherence loss in stroboscopic signal processing}\label{Modeling}

Let us represent the complex models of the received $\dot{x}(t)$  and the reference $\dot{a}(t)$  signals \cite{Zakharchenko15} including phase instability $\theta(t)$  in the form
\begin{equation}\label{xa}
  \dot{x}(t) = \sum_{k=0}^N {A(t-kT)    e^{j [\omega_0t+\theta(t)] }}, \qquad
  \dot{a}(t) = \sum_{k=0}^N {A_1(t-kT_1)e^{j\omega_1 t}},
\end{equation}
where $\dot{A}(t)$  and $\dot{A}_1(t)$ are the complex envelopes providing high range resolution; $T$, $T_1$ are the repetition periods of the signal and  the strobe; $T_1=T+\Delta T$;  $\Delta T=2TV_r/c$  is  the sampling increment ($\Delta T \ll T,T_1$).

The value of stroboscopic sample of a signal in the $k$-th sampling period can be presented as
\begin{equation}\label{yk}
  \dot{y}_k = \frac{1}{2T}\int_{kT}^{(k+1)T} A(t-kT)A_1(t-kT_1)e^{j[\Omega t+\theta(t)]} dt.
\end{equation}
Let us assume $\theta(t)$  be a stationary random zero-mean process with correlation window  $\tau_\theta$ exceeding the duration of strobe signals. Let us suppose also that slow phase displacements are tracked by stabilizing system, wherefore one can neglect the correlation of adjacent samples $\{\theta_k\}$  and set $\langle\theta_i\theta_k\rangle=\sigma^2_\theta\delta_{ik}$ . This allows for presentation of the sample $\dot{y}_k$  in the form
\begin{equation*}\label{yk2}
  \dot{y}_k \approx \frac{e^{j[\Omega kT_1+\theta_k]}}{2T} \int_{0}^{T} A(t^{\prime}) A_1(t^{\prime}-k\Delta T) dt = \dot{y}_{k0} e^{j\theta_k}.
\end{equation*}
where $\theta_k=\theta(kT_1)$  is the sample of the random process $\theta(t)$  and $\dot{y}_{k0}$  stands for the stroboscopic sample \eqref{yk} when there is no phase instability in the signal processing system. To ensure the mode of “ultra-high-resolution” of radar signatures of asteroids \cite{ZakhKovpatent14} one has to use nanosecond signals with pulse ratio of order $10^3-10^6$, thus, the aforementioned approximations are perfectly acceptable.

The average value (mathematical expectation) of samples \eqref{yk} obtained by averaging over phase $\theta_k$  can be expressed as
\begin{equation}\label{Myk}
  My_k = \langle \dot{y}_k\rangle = \beta \dot{y}_{k0},
\end{equation}
where $\beta=\langle \exp[j\theta_k] \rangle = \chi_{\theta}(1)$; $\chi_{\theta}(\nu)$  is the characteristic function of the distribution law of phase fluctuations $W(\theta_k)$. For the normal process with zero mean $W(\theta_k)=N(0,\sigma^2_{\theta})$ this value is equal to  $\beta= \exp[-0.5 \sigma^2_{\theta}] < 1$.

Variance  of samples $\dot{y}_k$  amounts to $D\dot{y}_k = \langle |\dot{y}_k-M\dot{y}_k|^2\rangle = |\dot{y}_{k0}|^2(1-\beta^2)$  under the assumptions made. The ratio of variance to squared  mean value is
\begin{equation}\label{eta}
  \eta = \frac{D\dot{y}_k}{|M\dot{y}_k|^2} = \frac{1-\beta^2}{\beta^2}
\end{equation}			
it has the meaning of relative power level of output noise resulting from phase fluctuations. This ratio can be significantly reduced by increasing the spectral transformation coefficient $N = T/\Delta T$  at the expense of sampling step $\Delta T$  decrease and by using data storage in a system digital filter. In this case the variance $D {y}_k$  will be lowered by a factor $m$ where $m$  is the accumulation coefficient \cite{Zakharchenko99}. Given that the filter and the spectrally compressed signal band are adaptively matched, the value $m$  is asymptotically equivalent to the number of sampling steps $\Delta T$   packed in the signal duration $\tau_x$: $m\sim \tau_x/\Delta T$.

\begin{figure}
  \includegraphics[width=1\textwidth]{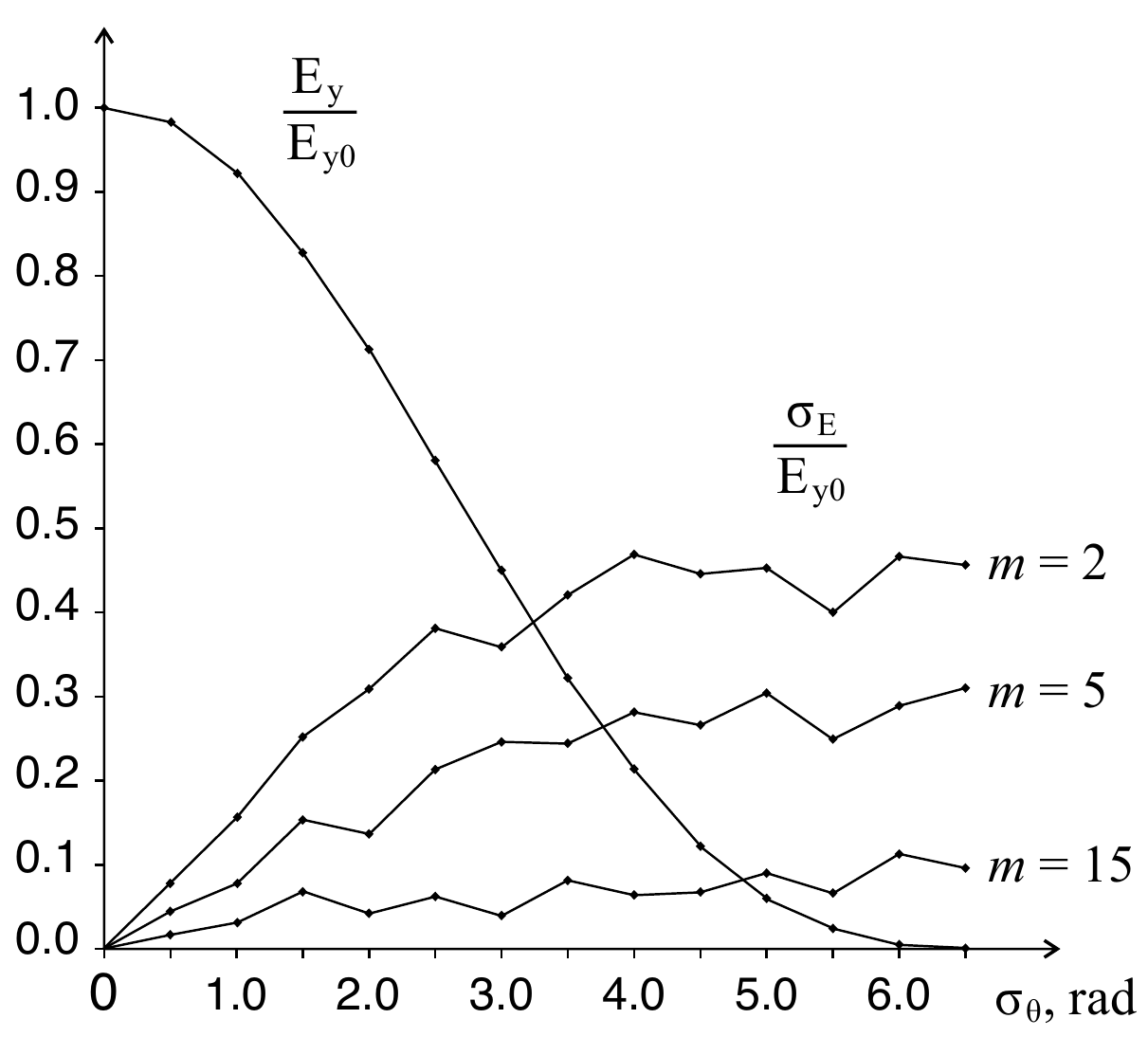}
\caption{Statistical characteristics of the output signal $y$ (mathematical expectation and root-mean-square deviation)  vs. phase instability $\sigma_{\theta}$  for different values of accumulation coefficient $m$.}
\label{fig:2}
\end{figure}

\section{Modeling results and quantitative estimates}\label{Estimates}

We performed numerical simulation of processing in the radio pulse strobing scheme (Fig.~\ref{fig:1}) of the signal $A(t)\cos[(\omega_0 +\Omega)t+\theta(t)]$  reflected by a single bright point of the surface of a moving asteroid. The envelopes of the probing and strobing $A(t)\cos \omega_0 t$  signals were chosen identical:  $A(t)= A_0\exp[-2(t/\tau)^2]$  with the effective duration $\tau$ determined according to the method of moments \cite{Gonorovskii86}: $\tau=2 ||tA(t)||/||A(t)||$. The random process $\theta(t)$  was specified as a sequence of uncorrelated samples $\theta_k=\theta(k T)$  with the normal distribution: $W(\theta_k)=N(0, \sigma^2_{\theta})$. The filter's transfer function  was rectangular  with the bandwidth equal to the width of the transformed signal power spectrum at $10\%$ $\max$ (–20dB) level.

Since the energy of received signal is important for  optimal reception under additive noise conditions \cite{Gonorovskii86}, the influence of phase instability was estimated as a decrease of the mean signal’s energy $E_y=||y(t)||^2$  at the output of the signal processing filter, relative to the energy $E_{y0}$  under full coherence condition ($\sigma_{\theta}=0$).

Fig.~\ref{fig:2} demonstrates statistical characteristics of the output signal  of the stroboscopic signal processing system for different accumulation coefficients in the filter obtained by statistical modeling at specified values  $\Omega=2\pi F$; $F=512$;  $\tau=0.015$; $N=2048$. The quantity $E_y/E_{y0}$   corresponding to signal’s energy decrease at the output of the band-pass filter of the stroboscopic signal processing system vs. phase instability is  plotted in Fig.~\ref{fig:2}. The relative errors caused by phase instability are also shown..  Presented results of statistical modeling are obtained by averaging over 100  simulation runs.

As previously noted \cite{Zakharchenko15}, for a cosmic object of about 50 m in size the range resolution of  $\delta r\sim 0.5$ m can be provided by coherent stroboscopic signal processing of signals with duration $\sim 3$ ns  in the X-range ($f_0\sim 10$ GHz) and frequency band  $\Delta f\sim 300$ MHz.

As it can be seen from Fig.~\ref{fig:2} decrease of mean received signal power by half corresponds to the phase instability of $\sigma_{\theta}\sim 3$ rad. The value of phase deviation $\theta$ caused by a short-time frequency instability $\Delta\omega$  and by finite signal propagation time $t_0=2R/c$  is
\begin{equation}\label{theta_ineq}
  \theta(t) \le \int_{t}^{t+t_0} {\Delta\omega(t^{\prime})dt^{\prime}}.
\end{equation}
It represents the Wiener process \cite{Kazakov73} with normal distribution. The upper-bound estimate of phase difference gives $|\Delta\theta|\le \max{|\Delta\omega(t^{\prime})|t_0}$.

For operation of stroboscopic radar station at range $R$ with acceptable phase instability of $\sigma_{\theta} <3$  rad it is required to ensure $|\Delta\omega| < c\sigma_{\Delta\theta}/2R$. This condition ensures that the noise level does not exceed $\sim 7$ dB at the accumulation coefficient $m>10$. At the contemporary technology level the stability of reference generators with relative error no grater than $\delta\sim 10^{-12}$, which corresponds to $\sim 0.01$ Hz in the X-range, is quite realizable \cite{Belov04}. Thus, the system can functionally operate within 5 million km distance.

\section{Conclusion}\label{Conclusion}

Loss of coherence in stroboscopic radar ranging systems  caused by phase instabilities of the reference sources leads to sensitivity degradation and is equivalent to the effects of modulating interference. Noise reduction at the output of the stroboscopic converter caused by  loss of coherence can be achieved by reducing the sample step $\Delta T=T/N$  with corresponding increase of the processing time.

\bigskip\medskip\noindent {\bf Acknowledgements}
\medskip

We are grateful to Vitaly Korolev for the help at vectorization of drawings and Victor Levi for careful reading of manuscript. The work is fulfilled within the framework of projects supported by grants from the Russian Foundation for Basic Research 15-47-02438-r-povolzhie-a and 14-02-97001-r-povolzhie-a.









\end{document}